\documentclass[12pt,a4paper]{article}
\usepackage[T2A]{fontenc}
\usepackage[cp866]{inputenc}
\usepackage[russian,english]{babel}

\usepackage{amssymb,amsmath}
\usepackage{graphicx}
\begin{document}

\begin{center}
{\Large 
Can Euclidean lattice quantum field theory be analytically\\[2mm] 
continued into Minkowski space? 
}\\[3mm]
{B.~P.~Kosyakov${}^{a}$, E.~Yu.~Popov${}^a$, and  M. A. Vronski{\u\i}${}^{a,b}$}\\[3mm]
{{\small ${}^a$Russian Federal Nuclear Center--VNIIEF, 
Sarov, 607188 Nizhni{\u\i} Novgorod Region, Russia;\\
${}^b$Sarov Institute of Physics {\&} Technology, Sarov, 607190 Nizhni{\u\i} Novgorod Region, 
Russia}\\} 
\end{center}
\begin{abstract}
\noindent{In this paper, we attempt to test whether Euclidean lattice quantum field theory can be analytically continued into Minkowski space via the inverse Wick rotation. 
Our discussion indicates that such an analytical continuation is impossible without first taking the lattice theory to the continuum limit. 
The obstacle is that discretization of spacetime converts local quantum field theory into a theory with a nonlocal form factor, for which the Wick rotation is infeasible.
}
\end{abstract}

\section{Introduction}
\label
{Introduction}
Lattice gauge quantum field theory,  originating from Wilson's famous work 
 \cite{Wilson}, is the subject of a large amount of textbook literature 
\cite{Seiler}--\cite{Knechtli}.
This is not so surprising if we take into account that 
this theory is currently the only tool for studying strongly coupled systems 
outside the framework of perturbative  quantum chromodynamics.
Impressive results in this area, obtained 
in numerical Monte Carlo simulations of supercomputers,
can be found in the above-mentioned books and in recent articles posted in the Cornell University electronic preprint archive under the hep-lat section.
However, there are also downsides to this successful venture. 
Here we would like to draw attention to the fact that there seems to be no direct analytic continuation of Euclidean lattice quantum field theory into Minkowski space ${\mathbb R}_{1,3}$, at least through the use of the famous Wick rotation \cite{Wick}.
As far as we know, this fact, despite its 
apparent clarity, has so far been overlooked.
Indeed, a widely held view, dating back to early lattice research, is that the transition matrix proposed by Wilson  \cite{Wilson} and studied in more detail by Creutz \cite{Creutz-}  allows one to relate the Euclidean lattice path integral of any gauge theory to its quantum mechanical description in Minkowski space.
Let us look at this issue more closely.

The mathematical basis of lattice quantum field theory is the Feynman path integral \cite{Feynman},  a rigorous mathematical definition of which is given in the formalism of imaginary time \cite{Kac}.
Therefore, all constructions in this theory begin with the discretization of Euclidean space ${\mathbb E}_4$ into a lattice with sites separated by distance $a$ and connected by links.
Matter fields, such as quark fields, are defined at lattice sites, whereas the gauge fields are defined on the links.
For processes 
produced by virtual particles, ${\mathbb E}_4$ is just the right arena. 
Many properties of virtual and real phenomena in the subnuclear realm coincide or are related to each other by obvious analytical continuation, so that to study them it is sufficient to operate in ${\mathbb E}_4$. 
However,  to be sure that all the results of calculations in lattice theory are relevant to the real world, one should check that the picture in ${\mathbb E}_4$ can be analytically continued to the picture in  ${\mathbb R}_{1,3}$.
The possibility of analytical continuation for time-ordered perturbation terms in {\it local} quantum field theory was first demonstrated by Wick \cite{Wick}, who considered the rotation of the time axis of integration by an angle of $\frac12\pi$ in the complex plane.
This 
form of analytic continuation formed the basis of subsequent more mathematically sophisticated constructions \cite{Schwinger}.
Thus, it is
reasonable to first 
pass to the limit $a\to 0$, thereby regaining the property of locality, and then make the transition to Minkowski space.
The strategy of Constructive Quantum Field Theory as it was pursued in the 70s to 80s  \cite{Brydges}, \cite{GlimmJaffe} is exactly that: first, find the continuum limit in the two-dimensional Euclidean spacetime, verify the Osterwalder--Schrader  axioms \cite{Osterwalder}, and then construct an analytical continuation to the Wightman theory.
The success of this strategy is due to the favorable facts that, in a two-dimensional world, the vacuum polarization is extremely weak (which, in technical language, is expressed as superrenormalizability, the absence of anomalies and phase transitions) and the presence of exactly solvable models.
By contrast, in the four-dimensional world, the vacuum polarization is large (that is, here one can only rely on renormalizability and a picture with a moderate number of theoretically controllable anomalies and phase transitions), and despite all efforts, exactly solvable models have not been found.
Therefore, it may be tempting to change the order of actions: first find an analytic continuation from ${\mathbb E}_4$ to  ${\mathbb R}_{1,3}$, and then move on to the continuum limit.
We will now show that these operations are non-commutative, because the discretization of spacetime introduces a form factor into the theory.

The interaction term of the field $\Phi$ with its source $J(\Phi)$ 
in form factor theories is typically given by
\begin{equation}
\int d^4\xi\,d^4\eta\,\Phi(\xi)K(\xi-\eta) J(\eta)\,,
\label
{form_factor-action}
\end{equation}  
where $K(\xi)$ is the form factor which serves as a cutoff at small spacetime distances.
The Fourier transform of this expression, 
\begin{equation}
\frac{1}{\left(2\pi\right)^4}\int d^4p\,{\tilde\Phi}(-p){\tilde K}(p){\tilde J}(p)\,,
\label
{form_factor-action-Fourier}
\end{equation}  
shows that ${\tilde K}(p)$  ensures ultraviolet convergence of scattering amplitudes in the Euclidean region provided that  ${\tilde K}(p)$ does not have pole singularities there and drops sufficiently rapidly as $p^2\to-\infty$.
It is well known that  ${\tilde K}(p)$ has such properties in {\it nonlocal} theories \cite{Meiman}--\cite{Efimov}.
The idea that the discretization of spacetime itself converts a local field theory into nonlocal one
has been around for quite a long time, in particular, in \cite{k07} it is argued that  the lattice formulation of gauge theories amounts to using the form factor
\[
K(x)=\delta(x^0-\ell_0)\,\delta(x^1-\ell_1)\,\delta(x^2-\ell_2)\,
\delta(x^3-\ell_3)
\]
\begin{equation}
=
\exp\left({\ell_0\frac{\partial}{\partial x^0}+\ell_1\frac{\partial}
{\partial x^1}+
\ell_2\frac{\partial}{\partial x^2}+\ell_3\frac{\partial}{\partial x^3}}\right)
\delta(x^0)\,\delta(x^1)\,\delta(x^2)\,\delta(x^3),
\label
{formf-QCD-lattice}
\end{equation}                          
whose Fourier transform is ${\widetilde K}(p)\propto\exp(i{\ell_\mu p^\mu})$, which is an entire function of the first order, in other words, we are dealing with a {nonlocal} form factor field theory in the sense of the Me{\u\i}man--Jaffe--Efimov classification
\cite{Meiman}--\cite{Efimov}.
However, the possibility of a Wick rotation in nonlocal form factor  theories is not guaranteed.
Below we show explicitly that  local lattice field theory is in fact a nonlocal form factor theory.
We restrict ourselves to considering  the simplest case of a two-dimensional theory of a noninteracting neutral scalar field (to avoid the irrelevant fermion doubling problem) and check whether there really is no direct analytic continuation from ${\mathbb E}_2$ to ${\mathbb R}_{1,1}$.

\section{ ${\mathbb E}_2$ vs ${\mathbb R}_{1,1}$}
\label
{E vs R}
In lattice theory, integration is replaced by summation, and differentiation is represented by a finite difference.
Both of these operations can be expressed as their continuous analogues using simple tricks.
To illustrate, we turn to the action of a free neutral scalar field theory
in two-dimensional Euclidean spacetime ${\mathbb E}_2$ discretized on a lattice,
\begin{equation}
S={a^2}\sum_n^\infty \sum_{\mu=1}^2 \left[\frac{1}{(2a)^2} \left(\phi_{n+{\hat\mu}}-\phi_{n-{\hat\mu}}\right)^2+m^2\phi_n^2\right].
\label
{quark-lattice-action}
\end{equation}                                         
Here the two-dimensional integration of the Lagrangian density function  is replaced by summation over the indices $n$ and ${\mu}$. 
The adoption of a symmetric discretization of the first derivative $\partial\phi/\partial x_\mu\to [\phi(x+a{\hat\mu})-\phi(x-a{\hat\mu})]/(2a)$, rather than a retarded $[\phi(x)-\phi(x-a{\hat\mu})]/a$  or an advanced $[\phi(x+a{\hat\mu})-\phi(x)]/a$  discretization, is due to the desire to preserve the self-adjoint property of the differentiation operator $i\partial/\partial x_\alpha$.

The joint presence of fields $\phi(\tau)$ and $\phi(\tau+a)$ in  (\ref{quark-lattice-action}) can be taken into account if we note that
\begin{equation}
\phi(\tau+a)=\sum_{k=0}^\infty \frac{a^k}{k!}\left(\frac{d}{d\tau}\right)^k \phi(\tau)=\exp\left( a\frac{d}{d\tau}\right)\phi(\tau)\,.
\label
{Newton} 
\end{equation}
This allows us to express discretized  differentiation by means of continuous techniques.
It is possible 
(and convenient)
 to work with Fourier transforms of such quantities,
\begin{equation} 
\phi(\tau+a)=\exp\left( a\frac{d}{d\tau}\right)\frac{1}{2\pi}\int_{-\infty}^\infty d\omega\,e^{i\omega \tau}{\tilde \phi}(\omega)=\frac{1}{2\pi}\int_{-\infty}^\infty d\omega\,e^{i\omega (\tau+a)}{\tilde \phi}(\omega)\,.
\label
{Schrdngr} 
\end{equation}
Then the Fourier transform of the product $\phi(\tau)\phi(\tau+a)$ has the form  ${\tilde\phi}(-\omega){\tilde\phi}(\omega)\exp(i\omega a)$.

Equation (\ref{quark-lattice-action}) contains two infinite summations, one of which is associated with a sequence of equally spaced points oriented along the ``time direction'' corresponding to the zero value of the $\mu$ index, and the other with a sequence of points in the perpendicular ``spatial direction'' corresponding to $\mu=1$.
We are interested here only in the summation along the ``time sequence''.
This operation can be transformed into integration over a continuous variable in infinite limits using the following  trick.
Consider a sequence whose terms ${\tilde Q}(\omega)_n$ sum from $n=0$ to $n=\infty$.
Expand $n$ to complex values $z$, which implies ${\tilde Q}(\omega)_n\to {\tilde Q}(\omega;z)$.
We next multiply each term ${\tilde Q}(\omega;z)$ by
\begin{equation}
\left(\frac{1}{2\pi i}\right) \frac{e^{i\pi z}}{\sin\pi z}\,,
\label{KG} 
\end{equation}
and integrate the result over a small closed contour counterclockwise in the $z$ plane around the point $z=n$. 
We combine all these contours into a single contour going above the real axis from $\infty$ to 0, capturing half of the contour around $z=0$, and then going under the real axis in the opposite direction to $\infty$.
For certain values of $\omega$ and other parameters included in  ${\tilde Q}(\omega;z)$,  the integration contour can be deformed to a straight line parallel to the imaginary $y$-axis, intersecting the real $x$-axis in the interval $(-1,0)$.
The sum of the series becomes the Mellin--Barnes integral of ${\tilde Q}(\omega;z)$ multiplied by  $\left(2\pi i\sin\pi z\right)^{-1}$, with
the convergence of this integral being  the same as the convergence of the original series.
\begin{center}
\begin{tabular}{ccc}
\includegraphics[height=0.175\textwidth]{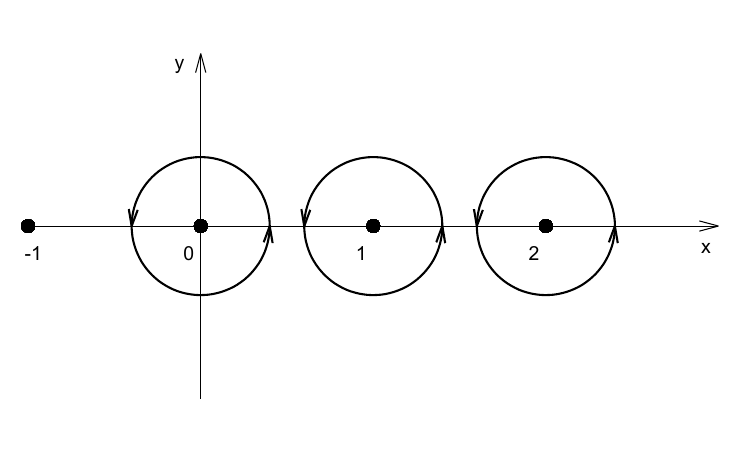}&
\includegraphics[height=0.175\textwidth]{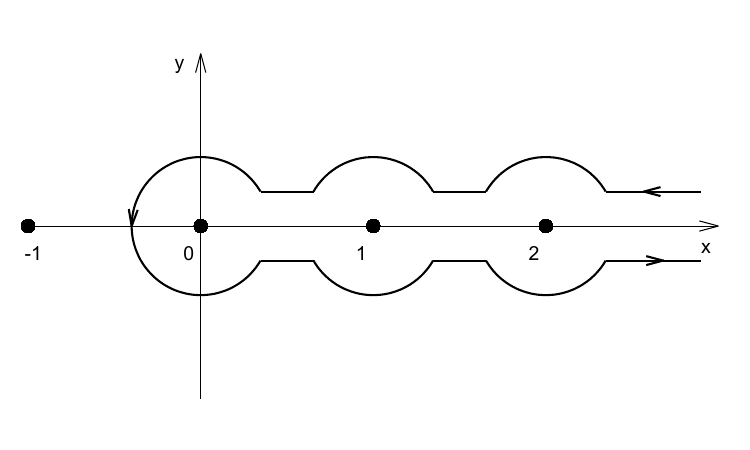}&
\includegraphics[height=0.175\textwidth]{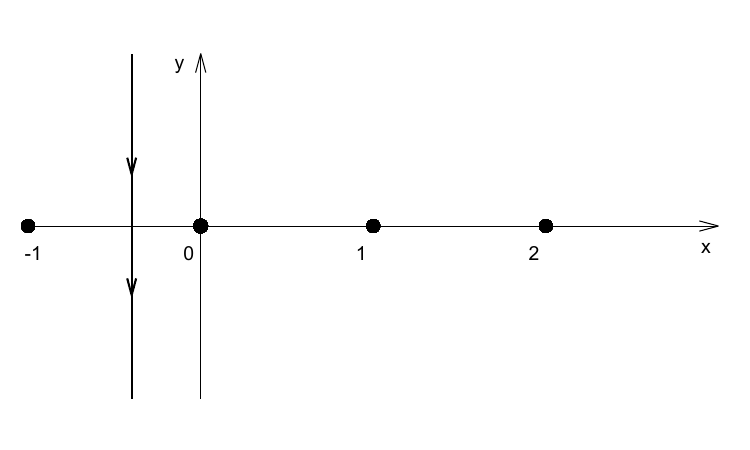}\\
\end{tabular}
\end{center}

Let us identify $\tau$ with $iy$, which is acceptable, since  $\tau$ in Eqs.~(\ref{Newton}) and  (\ref{Schrdngr}) was considered to be an arbitrarily chosen variable.
Now that the integration over $\omega$ is performed first, our concern remains to check whether an analytic continuation $\omega\to -ip^0$ is possible using the inverse Wick rotation of the integration axes by an angle of $\frac{\pi}{2}$. 
This would be truly achievable if the integrals over arcs of infinite radius in the first and fourth quadrants of the complex $\omega$ plane, connecting the ends of the integration axes, vanished.
Let us check if this is true.
We write  $\omega=\Omega e^{i\theta}=\Omega\left(\cos\theta+i\sin\theta\right)$ and consider the behavior of  $|e^{i\omega (\tau+a)}|$, which is
$e^{-\Omega (\tau+a)\sin\theta}$.
For $0<\theta<\frac{\pi}{2}$ we have  $\sin\theta\ge \frac{2\theta}{\pi}$, therefore 
 $|e^{i\omega (\tau+a)}|\le e^{-\Omega (\tau+a)\frac{2\theta}{\pi}}$.
In the interval  ${\pi}<\theta<\frac{3\pi}{2}$ 
the behavior of  $|e^{i\omega (\tau+a)}|$ is similar.  
If $\tau>0$, and moreover $\tau>|a|$, then the integral over an arc of infinite radius in the first quadrant of the complex domain is equal to 0.
With reference to (\ref{quark-lattice-action}), we must take into account the possibility of both positive and negative values of $a$.
However, for $\tau<-|a|$ this integral diverges.
It is thus clear that the analytical continuation of lattice field theory from ${\mathbb E}_2$ to ${\mathbb R}_{1,1}$ via the inverse Wick rotation is impossible.

This argument can be extended to the more general case of the Euclidean lattice theory of interacting fields in ${\mathbb E}_4$.

\section{Summary}
\label
{summary}

The simplest example of a free scalar field defined on a two-dimensional Euclidean lattice spacetime shows that an analytical continuation of this theory to a pseudo-Euclidean two-dimensional  lattice by means of an inverse Wick rotation is impossible. 
The reason for this result is simple and clear: the discretization of  spacetime  makes any local field theory a nonlocal form factor theory.
In more technical terms, the presence in the integrand of a product of fields that depend on both a given lattice site and neighboring sites entails the appearance in its Fourier transform 
an additional factor that grows exponentially in the ultraviolet region. Accordingly, the integral over an infinite-radius arc connecting the integration contours in the Euclidean and pseudo-Euclidean domains tends to infinity.

What are the implications of this result?
We would like to highlight one, which we believe is important both formally and conceptually. 
The lattice approach is based on the Feynman functional integral.
Its measure can be well defined only in Euclidean spacetime. 
If the results of integration in a lattice theory might be analytically continued to ${\mathbb R}_{1,3}$, then one could try to give a mathematical meaning to the measure of the Feynman functional integral in its original formulation. 
It seems that such a possibility is not feasible.

As for the conceptual consequences, they are less clear, since no correspondence has been established between  perturbative phenomena in conventional renormalizable field theories with small dimensionless coupling constants and their counterparts in lattice theories. Consider, for example, processes with timelike momenta in intermediate states, characterized by tree diagrams. Such processes are associated with the semiclassical regime. In the absence of an analytical continuation from ${\mathbb E}_{4}$ to ${\mathbb R}_{1,3}$, identifying such processes in lattice theory is hardly possible. Therefore, the very concept of a semiclassical regime is lost here.

\section*{Acknowledgments}
We thank Michael Creutz,  Alekse{\u\i} Morozov, and Erhard Seiler for fruitful discussions.

\section*{Conflict of interest statement}
The authors declare the absence of conflict of any competing interests.

\section*{Funding statement}
This work was funded by the  Russian Federal Nuclear Center--VNIIEF. 
No additional grants and sources of funding were received for conducting or supervising this study.

\end{document}